\documentclass[aps,nofootinbib,pra,amsmath,amssymb,twocolumn]{revtex4}
\usepackage{amssymb}
\usepackage{color}
\usepackage{graphicx}
\usepackage{epsfig,amssymb,amsmath}
\usepackage{chngcntr}
\usepackage[normalem]{ulem}

\usepackage{color}

\def\>{\rangle}
\def\<{\langle}
\def\d{\partial}
\def\k{\kappa}
\def\a{\alpha}
\def\vk{{\vec\kappa}}
\def\vr{{\vec r}}
\def\vrho{{\vec\rho}}
\def\comment#1{}
\def\labell#1{\label{#1}}
\def\togli#1{}

\begin{document}

\title{Gaussian beam quantum radar protocol} \author{Lorenzo
  Maccone$^1$, Yi Zheng$^{2,3}$ and Changliang Ren$^4$
}\affiliation{{\vbox{$1.$~Dip.~Fisica and INFN Sez.\ Pavia, University
      of Pavia, via Bassi 6, I-27100 Pavia, Italy}}\\\vbox{$2.$CAS Key
    Laboratory of Quantum Information, University of Science and
    Technology of China, Hefei 230026, People’s Republic of
    China}\\\vbox{$3.$ CAS Center for Excellence in Quantum
    Information and Quantum Physics, University of Science and
    Technology of China, Hefei 230026, People’s Republic of
    China}\\\vbox{$4.$ Key Laboratory of Low-Dimensional Quantum
    Structures and Quantum Control of Ministry of Education, Key
    Laboratory for Matter Microstructure and Function of Hunan
    Province, Department of Physics and Synergetic Innovation Center
    for Quantum Effects and Applications, Hunan Normal University,
    Changsha 410081, People’s Republic of China}}
\begin{abstract}
  We present an entangled quantum radar protocol. It consists in
  scanning the sky with a thin Gaussian beam and measuring the travel
  time of the radiation reflected from the target, as in conventional
  radars. Here the Gaussian beam is composed of $N$ photons entangled
  in the frequency degrees of freedom. We show that this provides a
  $\sqrt{N}$ quantum enhancement over the unentangled case, as is
  usual in quantum metrology.
\end{abstract}
\maketitle
In this paper we introduce a radar protocol that can achieve quantum
enhanced ranging and target detection. Other quantum radar protocols
\cite{others1,others2,others3,others4} are typically based on the
quantum illumination primitive \cite{qillum1,qillum2}: they can only
discriminate the target presence or absence {\em at a predetermined
  specific point}, which implies that one must scan the whole 3d space
in search of a target, which is impractical and
time-consuming. Recently, a protocol that claimed to achieve quantum
enhanced ranging capabilities was proposed in
\cite{qradar}. Unfortunately, its theoretical analysis was flawed as
it was based on an incorrect optical transfer function
\cite{erratum}. This paper then, to our knowledge, presents the only
known three dimensional quantum enhanced radar protocol that can give
quantum-enhanced ranging to the target: we remind that radar stands
for RAdio Detection And Ranging. A protocol for enhanced ranging in
the idealized one-dimensional case was presented in \cite{qps}, and
this protocol is a sort of 3d extension of it. The analysis presented
here is agnostic to the wavelength used, so the same protocol can be
used also in the optical regime (lidar). It is also more practical
than the previous protocol \cite{qradar} since it does not require
wideband radiation entangled in the wave vector $\vec k$ which would
require large antennas or antenna arrays: the protocol presented here
only employs Gaussian thin beams that can be produced with small
antennas (or lasers). The beams are entangled only in the frequency
degrees of freedom, which is more practical. There is no quantum
enhancement in the transversal direction, although this feature can be
added to our protocol, for example using the techniques presented in
\cite{fabre}, namely by injecting squeezed vacuum in the modes
orthogonal to the Gaussian mode used by the protocol, or with similar
techniques.

As in the case of most other quantum metrology protocols
\cite{review0,review1,review2,review3,review4}, we show an enhancement
in the precision of the order of $\sqrt{N}$, where $N$ is the number
of photons employed in the ranging procedure. Namely, we show that
this protocol can achieve the Heisenberg bound in precision in the
ideal noiseless situation. As usual, the situation becomes extremely
more complicated in the presence of noise, such as loss of photons,
but the usual general procedures and techniques to deal with noise can
be applied also in this case \cite{rafalguta,davidov,qec,dep},
e.g.~one can increase the robustness by reducing the entanglement (and
hence the precision gain) \cite{pranoise}.

The outline follows: we start in Sec.~\ref{s:transfer} by showing how
one can consistently derive the correct transfer functions in quantum
optics, based on the quantization of the electromagnetic (EM) field
which is reviewed (to set the notation) in App.~\ref{s:quant}. We then
show in Sec.~\ref{s:classical} how these techniques can be used to
give a quantum description of the usual classical radar
protocols. This is useful to show what are the ultimate bounds (due to
quantum mechanics) that can be achieved by these protocols in the
absence of entanglement. Finally, in Sec.~\ref{s:protocol} we
introduce and analyze our proposed protocol, and show its $\sqrt{N}$
enhancement.

\section{Quantization of optics through transfer functions}\labell{s:transfer}
In this section we review how optical transfer functions can be
consistently quantized. The notation and the framework we employ is
given in App.~\ref{s:quant}. The linearity and shift-independence form
of the Helmholtz equation \eqref{helmholtz} implies that its solutions
$U_\omega(\vec r)$ can be shifted in space, e.g.~along the $z$ axis
\cite{saleh}:
\begin{align}
U_\omega(\vec r\,')=\int d^2\vec r_{t}\:U_\omega(\vec r_{t},z)\;h_\omega(\vec r_{t}\,',\vec r_{t},z'-z)
\labell{transf}\;,
\end{align}
where the $t$ index represents the two-dimensional transverse vector
$\vr_{t}=(x,y)$, and where $\vec r=(\vec r_{t},z)$ and $h_\omega$ is
the transfer function that takes the solution $U_\omega$ at the $xy$
plane at position $z$ and, with a convolution, moves it to the $xy$
plane at position $z'$ (so the left-hand-side is independent of
$z$). This allows us to obtain the field at all positions starting
from the boundary values of the field in the plane $xy$ at position
$z$. Of course, the general solution of the field is given by
\eqref{aa1}, where one must sum $U_\omega$ over all components
$\omega$. Indeed, replacing Eq.~\eqref{transf} into \eqref{aa}, we
obtain the whole field $A({t},\vr\,')$ at position
$\vr\,'=(\vr\,'_{t},z')$ from a field on the plane $xy$ at position
$z$ at time ${t}=0$ (boundary conditions):
\begin{eqnarray}
A^+({t},\vec r\,')=\int d\omega\:d^2\vec r_{t} U_\omega(\vec r_{t},z)\:h_\omega(\vec r_{t}\,',\vec r_{t},d)e^{-i\omega {t}}
\labell{ltt},
\end{eqnarray}
where, for simplicity of notation, we consider only the
positive-energy component of the field $A^+$, namely only the first
term in the integral of \eqref{aa}.  

This transfer function formalism is developed in the classical case
but it can be transferred to the quantum case by first expressing the
solutions $U_\omega(\vec r)$ in terms of plane waves
$e^{i\vk\cdot\vec r}$, and then associating to each plane wave an
amplitude $a(\vk)$ as done in the customary EM quantization (see
App.~\ref{s:quant}). Namely,
\begin{eqnarray}
  A^+({t},\vec r\,')=\!\int\! d^2\vec r_{t}\:d^3\vk \:h_{\omega_\k}(\vec r_{t}\,',\vec r_{t},d)
  a(\vk)e^{-i(\omega_kt-\vk\cdot\vec r)}
\labell{quantumtransfer},
\end{eqnarray}
where $\vec r=(\vec r_{t},z)$, $\vec r\,'=(\vec r\,'_{t},z')$, and the
integral over $\omega$ is contained in the integral over $\vk$, since
$\omega_\k=c\k$. [More rigorously, the integral over $\omega$ comes
from \eqref{ltt}, whereas in the input field we are considering only
the $\omega$ component $U_\omega(\vr)$, so we need to integrate only
on the directions $\vk/\k$ as discussed below Eq.~\eqref{ss}.]  For
example, $\vec r$ may represent the object plane and $\vec r\,'$ the
image plane in an imaging apparatus, whose transfer function is given
by $h_\omega$ \cite{qimaging}. Eq.~\eqref{quantumtransfer} is the main
result of this section.

This is the field {\em operator}, so by itself it says nothing about
the physics: operators in quantum mechanics only acquire values when
applied to states, e.g.~the probability
$p({t},\vec r)\propto|\<0|A^+|\psi\>|^2$.  Alternatively, we may be
interested in other expectation values of the field in state
$|\psi\>$. The field degrees of freedom (including its boundary
conditions) are encoded into $|\psi\>$.  E.g., for a single photon
with $\psi(\vk)\propto\alpha(\vk)$ [see Eq.~\eqref{spac}], we
have\comment{Lorenzo's note pg z1}
\begin{align}
\nonumber&  \<0|A^+({t},\vr\,')|\psi\>\\&=\int d^3\vk\: d^3\vk'\:d^2\vr_{t}\: h_{\omega'}\:e^{-i(\omega' {t}-\vk'\cdot\vr)}\alpha(\vk)
\nonumber  \<0|  a(\vk')  a^\dag(\vk)|0\>\\
\nonumber  &=\int d^3\vk \:d^2\vr_{t} \:h_{\omega}\:e^{-i(\omega {t}-\vk\cdot\vr)}\:\alpha(\vk)
  \\&=\int d\omega\: d^2\vr_{t}\: h_{\omega}(\vr_{t}\,',\vr_{t},d)\:U_\omega(\vr)
\labell{esemp}\;,
\end{align}
where we used the fact that integrating $\alpha(\vk)e^{i\vk\cdot\vr}$
over the directions of $\vk$, one obtains $U_\omega(\vr)$ with
$\omega=c\k$, as is clear by the comparison between Eqs.~\eqref{aa}
and \eqref{ss}. This result is what one would expect from
\eqref{transf} by integrating over $\omega$ both members.
\subsection*{Free field transfer function}
The specific form of the function $h_\omega$ depends on what is
present between the two $xy$ planes at $z$ and $z'$, and on the
approximations used. In the case of vacuum propagation with the
Fresnel approximation, we get (\cite{saleh}, Eq.~4.1-14)
\begin{align}
  h_\omega(\vec r_{t}\,',\vec r_{t},d)=\frac {i\k}{2\pi d}e^{-i\k(\vec r_{t}-\vec r_{t}\,')^2/d}\:e^{-i\k d}
  \labell{vtrans}\;,
\end{align}
with $d=z'-z$ the distance between the two planes. While the
Rayleigh-Sommerfeld diffraction can give better results in some cases,
in the regimes we are interested, the Fresnel approximation that gives
rise to \eqref{vtrans} is sufficient to our aims.\togli{Instead, from
  Fraunhofer diffraction we obtain a slightly different, but more
  accurate transfer function
\begin{align}
  h_\omega(\vec r_{t}\,',\vec r_{t},z-z')\propto e^{i\vk\cdot\vec r_{t}}e^{-iC(\vec r_{t}\,'\cdot\vec r_{t}+d^2)}
\labell{rs}\;,
\end{align}
with $C=\k/d$.\comment{I took this from Eq.(3)
  of Yi Zheng's notes. I'm not sure where it comes from. Moreover
  there's a problem that the left-hand-side depends only on the
  modulus $|\vk|$, whereas the right-hand-side seems to depend on the
  vector $\vk$. It doesn't really matter: in the far field we're
  using, Eq.~\eqref{vtrans} is good enough}.} We will be using
Eq.~\eqref{vtrans} in the following.
\section{Quantum treatment of a classical radar/lidar protocol}\labell{s:classical}
A radar/lidar works by scanning the sky with a directional beam and
measuring the time it takes for it to be bounced back. The direction
of the beam and the time of flight suffice to do a full 3d
localization of the target. In this section we analyze a classical
radar/lidar protocol using quantized light to show what are the
ultimate bounds imposed by quantum mechanics to such classical
(unentangled) protocols.

As directional beam, we consider a Gaussian beam. For simplicity we
will consider the target as a perfectly (or partially) reflecting
mirror orthogonal to the beam direction, of size larger than the beam
waist at the target location. In this way we are guaranteed that the
beam that returns to the antenna is still in a (possibly attenuated)
Gaussian beam, see Fig.~\ref{f:protocol}. The case in which the target
is smaller than the beam waist should also not be too difficult: the
returning beam will be a spherical wave originating at the target.

\begin{figure}[hbt]
\begin{center}
\epsfxsize=.9\hsize\leavevmode\epsffile{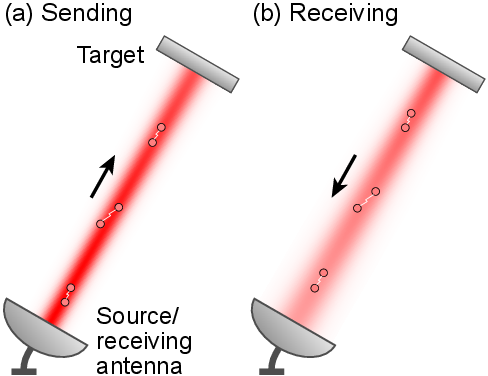}
\end{center}
\vspace{-.5cm}
\caption{Sketch of the quantum radar protocol. A Gaussian beam
  composed by frequency-entangled photons bounces off the target and
  returns to the sender's location. By measuring the average photon
  round-trip time, the sender can recover the target's position with
  quantum enhanced accuracy.} \labell{f:protocol}\end{figure}

The Gaussian beam for every frequency $\omega$ has an amplitude
$U_\omega(\vec r)=\varphi(\omega)G_\omega(\vr)$, where
$\varphi(\omega)$ is the spectral amplitude (the amplitude for each
frequency $\omega$ in the light) and $G_\omega$ is (\cite{saleh},
Eq.~3.1-7)
\begin{eqnarray}
  G_\omega(\vec r)\propto\frac1{W(z)}
  e^{-\frac{\k r_{t}^2}{2z_0W^2(z)}}e^{-i[\k z+\frac{\k r_{t}^2}{2zR(z)}-\arctan(z/z_0)]}
\labell{gaussianb}\;
\end{eqnarray}
where $W(z)\equiv\sqrt{1+z^2/z_0^2}$, $R(z)\equiv1+z_0^2/z^2$, $z_0$
is a (length) constant that, together with the direction of the $z$
axis, fully specifies $G_\omega(\vr)$. It is possible to check
\comment{Lorenzo's notes pg u} that this solution for fixed $z$ is
propagated to an arbitrary $z'$ through the transfer function
\eqref{vtrans}. The field $A$ is obtained by integrating $U_\omega$
over $\omega$, as in Eq.~\eqref{aa1}:
\begin{align}
A({t},\vr)=\int d\omega\:e^{-i\omega {t}}\:\varphi(\omega)\:G_\omega(\vr)
\labell{aas}\;.
\end{align}

We now consider a single photon in a Gaussian beam\footnote{It would
  be more appropriate to use a coherent state (or a thermal state) to
  model a classical beam, but since the photons in coherent states are
  completely uncorrelated (Poissonian statistics), one can easily
  obtain the same arrival statistics as a coherent state $|\alpha\>$
  (or its thermal mixtures) by considering what happens to
  $N=|\alpha|^2$ uncorrelated single photons. (Of course the photon
  number statistics will be different!) \comment{we could also
    calculate what happens on a coherent state, it should be easy.}}.
Since the light intensity $|A(\vec r)|^2$ at each point is directly
proportional to the probability of finding the photon there (as
discussed above) \cite{saleh,mandel}, we can choose
$\tilde\psi(\vec r)\propto A=\int d\omega\: \varphi(\omega)\:G_\omega(\vec
r)$, using \eqref{aas} with ${t}=0$ (because of the Heisenberg
picture) and the proportionality constant chosen by the normalization
condition. Namely, the photon wavepacket has probability amplitude
proportional to $G_\omega(\vr)$ for each frequency $\omega$, and the
probability amplitude of having frequency $\omega$ is given by
$\varphi(\omega)$. \comment{Lorenzo's notes pg y2.} So the state is
\begin{eqnarray}
  \labell{stat}  |\psi\>=\int d^3\vec\rho\:\tilde\psi(\vec\rho)\:a^\dag(\vec\rho)|0\>
  =
  \int d^3\vk'\:\tilde G(\vk')\,a^\dag(\vk')|0\>
\;,
\end{eqnarray}
where $\tilde G(\vk)$ is the Fourier transform of
$\varphi(\omega)\:G_\omega(\vec\rho)$, which clearly only contains the
frequency $\omega$ (the amplitude $\varphi$ is included in
$\tilde G$). We can write the field at the image plane, i.e.~the
detector position $\vr\,'$ in terms of the field at the target
position $\vr$ using \eqref{quantumtransfer}, with the transfer
function \eqref{vtrans}. Then, the probability amplitude of finding
the photon in ${t},\vr\,'$ is
\begin{eqnarray}
  \<0|A^+({t},\vr\,')|\psi\>=   \qquad\qquad\labell{avuo}\;&&\\
  \nonumber
\int d^2\vec r_{t}\:d^3\vk \:h_\omega(\vec r_{t}\,',\vec r_{t},d)\:
    e^{-i(\omega_kt-\vk\cdot\vec r)}    \<0|a(\vk) \times\qquad&&\\
  \nonumber
\int d^3\vk'\tilde G_{\omega'}(\vk')\:a^\dag(\vk')|0\>
    =\qquad&&\\  \nonumber
    \int d^2\vec r_{t}\:d^3\vk
    \:h_\omega(\vec r_{t}\,',\vec r_{t},d)\:e^{-i(\omega {t}-\vk\cdot\vr)}\:\tilde G(\vk)\propto&&\\
\nonumber
    \int d^2\vec r_{t}\:d\omega\:h_\omega(\vec r_{t}\,',\vec r_{t},d)
    \:e^{-i\omega {t}}\:\varphi(\omega)\:G_\omega(\vr_{t},z)
,&&
\end{eqnarray}
where we used the commutator \eqref{commut} in the second equality,
and in the third we used the far field condition
$\k_z\sim\k\gg\k_x,\k_y$ to separate the integral over $\vk$ into a
frequency and a transverse part $\vk_{t}$:
$d^3\vk\propto d\omega d^2\vk_{t}$, so that the integral of
$e^{i\vk\cdot\vr}\psi(\vk)$ over the transverse part of $\vk$ gives
the spatial field at frequency $\omega$, namely
$U_\omega(\vr)=\varphi(\omega)G_\omega(\vr_{t},z)$ with
$\vr=(\vr_{t},z)$ [compare Eqs.~\eqref{aa} and
\eqref{ss}]. Eq.~\eqref{avuo} is compatible with what one would expect
from the transfer function of the classical amplitudes: see
Eq.~\eqref{transf} when the time evolution of the output field is
added and both members are integrated over $\omega$.  We now use the
fact that the free space transfer function \eqref{vtrans} applied to a
Gaussian beam \eqref{gaussianb} translates it forward by a factor $d$,
the distance between target and receiver (namely, a Gaussian beam is
transformed in a Gaussian beam thanks to the hypothesis that the
target is a partially reflecting mirror larger than the beam
waist). Then, \eqref{avuo} becomes
\begin{align}
\<0|A^+({t},\vr\,')|\psi\>=    \int d\omega
    \:e^{-i\omega {t}}\varphi(\omega)G_\omega(\vr_{t}\,',z+d)
\labell{fi}\;.
\end{align}
As expected, at the image plane at position $z'$, it gives a
pulse that is delayed by the transit time to the target. To see this,
consider the expression \eqref{fi} at the center of the image plane
$\vr_{t}\,'=0$ where, from \eqref{gaussianb} we see that
$G_\omega(\vr_{t}\,'=0,z)\propto e^{-i[\k z-\arctan(z/z_0)]}$, so that \eqref{fi} becomes
\begin{align}
& \nonumber \int d\omega\:
  \:e^{-i\omega [{t}+(z+d)/c]-i\arctan((z+d)/z_0)}\varphi(\omega)
\\&  =\tilde\varphi({t}+(z+d)/c)\:e^{-i\arctan((z+d)/z_0)}
  \labell{llll}\;,
\end{align}
where $\tilde\varphi$ is the Fourier transform of
$\varphi$. Eq.~\eqref{llll} describes a pulse of spectral amplitude
$\varphi(\omega)$ and temporal amplitude $\tilde\varphi({t})$ that is
delayed by an amount $(z+d)/c$, where $d$ is the distance between target
and receiver and $z=d$ is the position of the target. By measuring the
time of arrival of the photon, one can obtain twice the distance $2d$
to the target, as expected for a radar. The statistical error in this
measurement is given by the width $\Delta\tau$ of
$\tilde\varphi(\tau)$, proportional to the inverse of the bandwidth
$\Delta\omega$ of $\varphi(\omega)$.

Now we could do the same calculation with a coherent state $|\alpha\>$
instead of a single photon state \eqref{stat}, with
$|\alpha\>=\bigotimes_\vk|\alpha(\vk)\>$ with $|\alpha(\vk)\>$
eigenstates of $a(\vk)$:
$a(\vk)|\alpha(\vk)\>=\alpha(\vk)|\alpha(\vk)\>$. This calculation
should give exactly the same outcome as a classical field amplitude
$\alpha(\vk)$, see Eq.~\eqref{ss}.
\section{Quantum radar/lidar protocol}\labell{s:protocol}
We now show how one can obtain an increased localization precision by
using frequency-entangled light.  For simplicity of notation, we will
consider only the case of $N=2$ entangled photons. This can then be
extended to arbitrary $N$.

For the $N$-photon state $|\psi_N\>$ of \eqref{stgen}, the probability
of detecting them at ${t}_1,\vr_1,\cdots,{t}_N,\vr_N$ is \cite{mandel}
\begin{align}
  p\propto  |\<0|A^+({t}_1,\vr_1)\cdots A^+({t}_N,\vr_N)|\psi_N\>|^2
  \labell{probn}\;.
\end{align}

Consider the biphoton entangled state with wavefunction
\comment{Lorenzo's note pg.aa}
\begin{align}
\tilde\psi_2(\vr,\vec\rho)\propto\int d\omega\:\varphi(\omega)\:G_\omega(\vr,\vrho)
\labell{bifot}\;,
\end{align}
which gives the probability amplitude of finding the two photons at
positions $\vr=(\vr_{t},z)$ and $\vrho=(\vrho_{t},\rho_z)$ (in the
Heisenberg picture there is no time evolution), and where
\begin{align}
  &\labell{gib}\;  G_\omega(\vr,\vrho)\equiv\\&\nonumber
  \tfrac1{W(z+\rho_z)}
  e^{-\frac{\k(\vr_{t}^2+\vrho_{t}^2)}{2z_0W^2}}e^{-i[\k(z+\rho_z)+\frac{\k(\vr_{t}^2+\vrho_{t}^2)}{2zR}-\arctan\frac{z+\rho_z}{z_0}]},
\end{align}
which represents two photons of identical frequency $\omega$ in a
Gaussian beam, see Eq.~\eqref{gaussianb}. Except for the
multiplicative term $1/W$ and the arctan term, $G_\omega$ is basically
a product of two Gaussian beam single-photon amplitudes. So we can
reuse the calculations above for the single-photon amplitude to find
that the temporal amplitude at the center of the image plane
$\vr_{t}=\vrho_{t}=0$ at the image plane position $z=\rho_z=z'$ is
given by the analogous of \eqref{llll}:
\begin{align}
\<0|A^+({t}_1,\vr)A^+({t}_2,\vrho)|\psi\>=\tilde\varphi({t}_1+{t}_2+2(z+d)/c)\:e^{i\theta}
\labell{am}\;,
\end{align}
with $\theta$ some irrelevant phase factor. From this, it is clear
that the time of arrival sum ${t}_1+{t}_2$ has an uncertainty
$\Delta\tau$,the width of $\tilde\varphi$. Which means that the
average time of arrival $({t}_1+{t}_2)/2$ is estimated to be the
correct value $d+z=2d$ with a statistical error $\Delta\tau/2$.
Instead, from \eqref{llll} we saw that, using a single photon state,
one estimates the time of arrival with an uncertainty $\Delta\tau$, so
the average time of arrival of two photons will be estimated with an
uncertainty $\simeq\Delta\tau/\sqrt{2}$. The $\sqrt{2}$ enhancement in
precision is the $\sqrt{N}$ gain that one expects from entanglement in
quantum metrology.

The biphoton analysis done here can be straightforwardly extended to
the case of $N$ entangled photons in a Gaussian beam.\comment{Check
  this: I didn't do it!} Namely, a state with wavefunction
\begin{align}
\psi(\vr_1,\cdots,\vr_N)\propto\int d\omega\varphi(\omega)G_\omega(\vr_1,\cdots,\vr_N)
\labell{nfot}\;,
\end{align}
where $G_\omega$ is a trivial generalization of \eqref{gib}. It gives
a $\sqrt{N}$ enhancement in the average photon time of arrival, which
translates into a $\sqrt{N}$ precision enhancement in the longitudinal
localization for each point in the sky scanned by the $N$-photon
Gaussian beam \eqref{nfot}, when one measures the average arrival time
$\sum_it_i/N$.

\section{Conclusions}\labell{s:conclusions}
In conclusion we have presented a quantum radar protocol that uses
entanglement in the frequency/wavelength degrees of freedom to provide
an quantum enhancement equal to the square root $\sqrt{N}$ of the
number $N$ of entangled photons employed. We have shown in detail how
the optical transfer function formalism can be employed in the fully
quantum regime we analyze here.
\appendix\section{Quantization of the EM field}\labell{s:quant}
In this appendix we review the usual theory for the quantization of
the electromagnetic field. This is useful to set the notation we use
in the paper, and also to keep track of the specific roles that all
the radiation degrees of freedom play in our protocol. Specifically,
it is useful to understand the peculiar role of the frequency degree
of freedom of the radiation that our protocol hinges on.
\subsection{Classical EM in the Coulomb gauge}
Start from the Maxwell equations in vacuum in the Coulomb gauge for
the scalar and vector potentials $\Phi$ and $\vec A$:
$\nabla^2\Phi({t},\vr)=0$,
$\square \vec A({t},\vec r)=\tfrac\d{\d t}\vec\nabla\Phi$, where
$\vec r=(x,y,z)$ is the spatial position, and
$\square=\nabla^2-\frac1{c^2}\frac{\d^2}{\d {t}^2}$ is the
d'Alamabertian with
$\vec\nabla=(\frac\d{\d x},\frac\d{\d y},\frac\d{\d z})$ in Cartesian
coordinates. Conventionally, since we are interested only in the
quantization of the electromagnetic waves, one chooses the specific
solution $\Phi=0$, which implies $\square \vec A({t},\vec r)=0$. For
simplicity of notation we will consider scalar fields $A({t},\vr)$ from now on:
the vectorial part can be added by introducing two independent
components, connected to the two polarizations of the em field (there
are two polarizations because, in the Coulomb gauge, the potential
$\vec A$ is transverse: $\vec\nabla\cdot\vec A=0$). We separate the
temporal and spatial degrees of freedom by taking a Fourier transform
over time:
\begin{align}
  A({t},\vec r)=\int_{-\infty}^{+\infty} d\omega\: e^{-i\omega t}\:U_\omega(\vec r)
  \labell{aa1}\;,
\end{align}
where $U_\omega$ is the component at frequency $\omega$. Since $A$ is
real, we must have $U_{-\omega}=U^*_\omega$, this condition can be
enforced automatically if we separate the integral into a sum of two
and change variable in the second:
\begin{align}
  A({t},\vec r)=\int_0^\infty d\omega[e^{-i\omega t}U_\omega(\vec r)+e^{+i\omega t}U^*_\omega(\vec r)]
  \labell{aa}\;.
\end{align}
For each component at frequency $\omega$, it is clear from \eqref{aa1}
that $\square A=0$ becomes the Helmholtz equation
\begin{align}
\nabla^2U_\omega(\vec r)=\frac{\omega^2}{c^2}U_\omega(\vec r).
\labell{helmholtz}\;
\end{align}
A convenient\footnote{Because of the linearity of the Helmholtz
  equation, any solution $U_\omega(\vec r)$ can be expressed as a sum
  of plane waves
  $U_\omega(\vec r)=\int d^3\kappa \:\a(\vk)\:e^{i\vec\kappa\cdot\vec
    r}$.} solution is in terms of plane waves
$U_\omega(\vec r)=\a(\vec\kappa) e^{\pm i\vec\kappa\cdot\vec r}$. The
real and imaginary part (or the modulus and phase) of $\a$ are the two
integration constants, and the wave direction $\vk/\k$ parametrizes
all the solutions, while we must choose $|\vec\kappa|$ such that
$\omega=\vec c\cdot\vec\kappa=\pm|\vec\kappa|c$. The sign parametrizes
two classes of solutions: we must choose $+$ if the wave vector
$\vec\kappa$ is parallel to the wave velocity $\vec c$, or $-$ if it
is antiparallel. The first case refers to the retarded waves, the
second to the advanced waves (\cite{jackson}, sec. 6.4). We usually
choose past boundary conditions, so we only use retarded
waves\footnote{If we were to choose future boundary conditions, we
  would need to consider only advanced waves, and if we were to choose
  mixed future-and-past boundary conditions \cite{einstein1909}, we
  would have to keep both solutions, which, incidentally, is a real
  problem in quantum field theory, as this leads to a non-Hamiltonian
  evolution of the electromagnetic field! Advanced waves can be seen
  as propagating negative energy in the forward time direction or
  positive energy in the negative time direction.}.  Summarizing, the
vacuum solution of the Maxwell equations is of the form
\begin{align}
  A({t},\vec r)=\int_{{\mathbb R}^3} d^3\vec\kappa[\a(\vec k)\:e^{-i(\omega_\kappa
  t-\vec\kappa\cdot\vec r)}+\a^*(\vec k)\:e^{i(\omega_\kappa
  t-\vec\kappa\cdot\vec r)}] \labell{ss}\;,
\end{align}
where $\a(\vec\kappa)$ is the positive-frequency
amplitude\footnote{Note that the Maxwell equations are solved also by
  negative frequency plane waves of the type
  $\a_-(\vec\kappa)\:e^{-i(\omega_\kappa t+\vec\kappa\cdot\vec r)}$
  but, as discussed above, we will ignore these solutions, by choosing
  past boundary conditions as is done usually.} and
$\omega_\kappa=|\vec\kappa|c$, so that the integral over $\vk$ takes
care of the integral over $\omega$ in \eqref{aa} (its modulus) and of
the integral over the directions $\vk/\k$ that enumerate all plane
waves.

\subsection{Quantum em: observables}
The energy of the electromagnetic field is
$H=\frac {\epsilon_0}2\int d^3\vec r[E^2({t},\vec r)+c^2B^2({t},\vec
r)]$, where $\vec E$ and $\vec B$ are the electric and magnetic
fields. In terms of the amplitudes $\a(\vec\kappa)$, one can show
that, in the Coulomb gauge\togli{\footnote{This is not true anymore in the
  Lorenz gauge, which make a covariant quantization of the em field
  rather problematic.}}, the energy is
\begin{align}
H=\frac 12\int d^3\vec\kappa\:(P^2_{\vec\kappa}+\omega_\kappa^2X^2_{\vec\kappa})
\labell{hamilt}\;,
\end{align}
with $P_\vk\propto i(\a_\vk^*-\a_\vk)$ and
$X_\vk\propto(\a_\vk^*+\a_\vk)$. Eq.~\eqref{hamilt} is the energy of a
collection of independent (noninteracting) harmonic oscillators (one
for each value of $\vec\kappa$), so we can quantize by considering $X$
and $P$ as ``position'' and ``momentum'' operators, promoting the
amplitudes $\a$ to operators $a$. Namely, we impose
$[X_{\vec\kappa},P_{\vec\kappa'}]=i\delta(\vec\kappa-\vec\kappa')$,
where the delta shows that they are independent oscillators for each
$\vec\kappa$. From the definitions of $X$ and $P$, this implies the
commutators
\begin{align}
[a(\vec\kappa),a^\dag(\vec\kappa')]=\delta(\vec\kappa-\vec\kappa')\;, [a(\vec\kappa),a(\vec\kappa')]=0
\labell{commut}\;.
\end{align}
The quantization of the general solution of the Maxwell equations in
the Coulomb gauge $\square A=0$, is then
\begin{align}
  A({t},\vec r)=\int d^3\vec\kappa\:[a(\vec k)\:e^{-i(\omega_\kappa
  t-\vec\kappa\cdot\vec r)}+a^\dag(\vec k)\:e^{i(\omega_\kappa
  t-\vec\kappa\cdot\vec r)}] \labell{campoq}\;,
\end{align}
namely Eq.~\eqref{ss} quantized. Importantly, since we are
introducing the time evolution in the operators, we are working in the
Heisenberg picture (or in the interaction picture with the free-field
Hamiltonian to evolve the operators). We are working with the vector
potential field $A$, but the electric field and magnetic fields are
trivially obtained from it:
$\vec E=-\frac\d{\d t}\vec A-\vec\nabla\Phi$,
$\vec B=\vec\nabla\times\vec A$, which give expressions very similar
to \eqref{campoq}, except for the fact that the derivatives introduce
a minus sign between the two terms of the right-hand-side (as
$\vec A$, also $\vec E$ and $\vec B$ have two independent components
since they are transverse: $\vec\nabla\cdot\vec B=0$ and, in vacuum,
$\vec\nabla\cdot\vec E=0$).

The intensity of the field is proportional to the time averaged square
$\<E^2\>_t$. This is basically equal to the average photon number in
the field. Indeed, from \eqref{ss} and the fact that
$\vec E=-\tfrac\d{\d t}\vec A$ we have that, classically,
$E^2=-\omega(\a^2e^{i\phi}+(\a^*)^2e^{-i\phi}-2|\a|^2)$ for a
classical field with only a single $\vk$, with
$\Phi=\omega {t}-\vk\cdot\vr$. This almost matches the quantum result
one would get for a coherent state $|\a\>$ for which $a|\a\>=\a|\a\>$
(which represents a classical field). A coherent state has
$E^2=-\omega(\a^2e^{i\phi}+(\a^*)^2e^{-i\phi}-2|\a|^2-1)$, where the
-1 term comes from the commutator $[a,a^\dag]=1$ (valid for the
quantization of a field with single $\vk$ vector). The time average
removes the terms with the phase, leaving only the average photon
number for a coherent state, i.e.~$\<\a|a^\dag a|\a\>=|\a|^2$. So
while $E^2$ does not coincide with the average photon number, the
time-averaged $E^2$ essentially does for classical fields.  Similar
considerations apply also to states with fixed photon number we
consider below, where $\<a^2\>=0$.\comment{check this: the discussion
  is valid for fields with single $\vk$ does it generalizes to
  arbitrary fields? I didn't check. I'm not even sure it's worth
  mentioning this: we don't need it}

\subsection{Quantum em: states}
In the classical case, we can choose a specific form of the
$\a(\vec\kappa)$ to obtain a specific solution of the Maxwell
equations (which can be done by choosing appropriate boundary
conditions for the field). In the quantum case, the
$\a\to a(\vec\kappa)$ are operators. There are two ways to assign a
value to them: (i)~have them act on eigenstates of the field (which
implies that the field is in a state where there are no quantum
fluctuations of the field). From the form of Eq.~\eqref{campoq}, it is
clear that the eigenstates of the field are quadrature eigenstates for
each $\vk$, where the quadrature is
$Q_\varphi\equiv(a\:e^{-i\varphi}+ a^\dag
e^{i\varphi})/\sqrt{2}$. These eigenstates are unphysical as they are
infinitely squeezed states with infinite average energy
$\hbar\omega_\k\<a^\dag(\kappa)a(\k)\>$. (ii)~we can calculate the
field expectation value on an arbitrary state $|\psi\>$ of the field
(which implies that we can calculate the average field, because there
are quantum fluctuations: measuring the field multiple times, we would
get different results).

How do we choose $|\psi\>$? It is the state of the em degrees of
freedom. Since these are given by a harmonic oscillator for each
$\vk$, the Hilbert space is a Fock space for each $\vk$, so the most
general state is given by
\begin{align}
  &\labell{stgen}  |\psi\>={\textstyle\sum_{n}}\gamma_n|\psi_n\>,\mbox{ with }\\&\nonumber
  |\psi_n\>=  \!\int\!\! d^3\vk_1\cdots d^3\vk_n\:\psi_n(\vk_1,\cdots ,\vk_n)a^\dag(\vk_1)\cdots a^\dag(\vk_n)|0\>,
\end{align}
where $\gamma_n$ is the probability amplitude to have $n$ photons,
$\psi_n$ is the joint probability amplitude that they are in modes
$\vk_1,\cdots,\vk_n$, and $|0\>$ is the vacuum. In this paper we only
consider states with a single component $\gamma_N=1$. In the
Heisenberg picture, the state does not evolve in time. The
 wavefunction normalization condition is
\begin{align}
\int d^3\vk_1\cdots d^3\vk_n\:|\psi_n(\vk_1,\cdots,\vk_n)|^2=1\ \forall n
\labell{norm}\;.
\end{align}
More rigorously, the state \eqref{stgen} refers only to the situation
in which all the photons are in different modes (namely, the $\psi_n$
does not contain Dirac $\delta$s over the $\vk_i$). The most general
situation is to have the $n$ photons distributed into $m\leqslant n$
modes ($m=1$ if all the $n$ photons are in one mode, $m=n$ if there is
one photon per mode, as above). The indistinguishable nature of the
photons implies that only $m$ of the
$\textstyle\left(\begin{matrix} m\cr n\end{matrix}\right)$
possibilities can be tracked, namely we can only know that $n_i$
photons are in mode $\vk_i$ for $i=1,\cdots,m$. In this case, the
wavefunction is
$\psi_n(n_1,\vk_1,\cdots,n_m,\vk_m)/\sqrt{n_1!\cdots n_m!}$ with
$\sum_in_i=n$, where the factorials appear because the $n_i$ photon
Fock state in a mode is given by
$|n_i\>=(a^\dag)^{n_i}|0\>/\sqrt{n_i!}$ and where $\psi_n$ is the
joint probability amplitude that the $n$ photons are partitioned as
$\{n_i\}$ {\em and} that their wave vectors are
$\{\vk_i\}$\togli{\footnote{In a sense, the $n_i!$ terms remind us that in a
  state $\int d^3\vk\:\psi_n(n,\vk)[a^\dag(\vk)]^n|0\>/\sqrt{n!}$ we
  are constraining all the differences in of the photon wave vectors
  $\vk_i-\vk_j=0$ for all $i,j$, whereas the sums are
  unconstrained. If we look at the same state in the position space,
  $\int d^3\vr_1\:\cdots
  d^3\vr_n\:\tilde\psi_n(n,\vr_1+\cdots+\vr_n)a^\dag(\vr_1)\cdots
  a^\dag(\vr_n)|0\>/\sqrt{n!}$, only the sum of positions is given,
  whereas all the differences $\vr_i-\vr_j$ are
  unconstrained.\comment{This footnote is not very convincing, I
    should do better, but it's not very important for this paper} }}.

For single photon states, only the term $n=1$ of \eqref{stgen}
survives. The spatial dependence of the wavefunction can be obtained
by taking the Fourier transform $\tilde\psi$ of $\psi\equiv\psi_1$:
\begin{align}
\int d^3\vk\:\psi(\vk)\:a^\dag(\vk)|0\>=\int d^3\vr\:\tilde\psi(\vr)\:a^\dag(\vr)|0\>
\labell{spac}\;,
\end{align}
where $a(\vr)\propto\int d^3\vk\:a(\vk)\:e^{i\vk\cdot\vr}$ is the
annihilator of a photon at position $\vr$, so that $\tilde\psi(\vr)$
is the probability that the photon is in $\vr$. (Note that, except in
the limit discussed in the next subsection, this is {\em not} in
general equal to the probability amplitude of measuring the photon at
position $\vr$, since there is a difference between the position of
the photon and of its energy, a well known problem in quantum field
theory, e.g.~\cite{posphoton1,posphoton2,mandel}. Indeed, as is clear from
the above analysis, the photon is obtained from the quantization of
the vector potential $A$, which is a gauge-dependent quantity, whereas
its energy is, clearly, a gauge-independent quantity.)

We can choose $\psi(\vk)={\cal N}\alpha(\vk)$, where $\alpha(\vk)$ is
the Fourier transform of the classical solution $A({t},\vr)$ of
\eqref{ss} and $\cal N$ is a normalization for
Eq.~\eqref{norm}. Indeed, as discussed below, $|A({t},\vr)|^2$ is the
light intensity at position ${t},\vr$, so it is proportional to the
probability of finding the photon at such position, so $A$ is the
probability amplitude, and its Fourier transform $\alpha(\vk)$ is the
probability amplitude in the $\vk$ space.

\subsection{Quantum em: photodetection}
It can be shown that for photodetectors with efficiency $\eta$,
sufficiently small temporal resolution $\tau$, and spatial resolution
$\sigma$, the probability of a photodetection at spacetime position
$({t},\vec r)$ is \cite{mandel}
$p({t},\vec r)\propto\eta\tau\sigma\<\psi|[A^+({t},\vec r)]^\dag
A^+({t},\vec r)|\psi\>$. In the case in which the system state
$|\psi\>$ contains a single photon, we can use the fact that $a$ is
the photon annihilator to simplify it to
$p\propto|\<0|A^+|\psi\>|^2$. To show this, consider
$|\psi\>=\int d^3\vk'\psi(\vk')a^\dag(\vk')|0\>$, with $\psi(\vk')$
the probability amplitude that the photon has wave vector $\vk'$ (so
that its Fourier transform can be interpreted as the probability
amplitude that the photon is in position $\vec r$). Then
\begin{align}
\nonumber&  A^+|\psi\>=\int d^3\vk\:d^3\vk'\psi(\vk')a(\vk)e^{-i(\omega_\k {t}-\vk\cdot\vec r)}
  a^\dag(\vk')|0\>=\\&\tilde\psi(\vec r-\vec ct)|0\>
\labell{der}\;,
\end{align}
where $\vec c$ is the speed of light with the direction $\vk/\k$ of the
beam. Eq.~\eqref{der} follows from the commutator \eqref{commut} and
the fact that $a|0\>=0$, and where $\tilde\psi$ is the Fourier
transform of $\psi$. [Note the use of the Heisenberg picture: the time
evolution is only in the operator $A^+$, not in the state.]

\section*{Acknowledgments}
This work received support from EU H2020 QuantERA ERA-NET Cofund in
Quantum Technologies, Quantum Information and Communication with
High-dimensional Encoding (QuICHE) under Grant Agree- ment 731473 and
101017733, from the U.S. Department of Energy, Office of Science,
National Quantum Information Science Research Centers, Superconducting
Quantum Materials and Systems Center (SQMS) under Contract
No. DE-AC02-07CH11359. L.M. acknowledges support from the PNRR MUR
Project PE0000023-NQSTI and from the National Research Centre for HPC,
Big Data and Quantum Computing, PNRR MUR Project CN0000013-ICSC.
Y.Z. thanks Jin-Shi Xu for the discussions. C.R. was supported by the
National Natural Science Foundation of China (Grant No.12075245,
12247105), the Natural Science Foundation of Hunan Province
(2021JJ10033).

\end{document}